\documentclass[preprint,showpacs,amsmath,amssymb,floatfix,pre]{revtex4-1}
\usepackage{graphicx}
\begin{document}

\title{Qualitative aspects of the phase diagram of J1-J2 model on the cubic lattice}

\author{Octavio D. R. Salmon $^{1}$}
\thanks{octaviors@gmail.com}

\author{Nuno Crokidakis $^{2}$}
\thanks{nuno.crokidakis@fis.puc-rio.br}

\author{Minos A. Neto $^{1}$}

\author{Igor T. Padilha $^{1}$}

\author{J. Roberto Viana $^{1}$}

\author{J. Ricardo de Sousa $^{1,3}$}

\affiliation{$^{1}$ Departamento de F\'{\i}sica, Universidade Federal do Amazonas, 3000, Japiim, 69077-000, Manaus-AM, Brazil \\
$^{2}$ Departamento de F\'{\i}sica, PUC-Rio, 
Rua Marqu\^{e}s de S\~{a}o Vicente 225 22451-900 Rio de Janeiro - RJ, Brazil \\
$^{3}$ National Institute of Science and Technology for Complex Systems, 3000, Japiim, 69077-000, Manaus-AM, Brazil}

\date{\today}

\begin{abstract}
\noindent
The qualitative aspects of the phase diagram of the Ising model on the cubic lattice, with ferromagnetic nearest-neighbor interactions ($J_{1}$) and antiferromagnetic next-nearest-neighbor couplings ($J_{2}$) are analyzed in the plane temperature versus $\alpha$, where $\alpha=J_{2}/|J_{1}|$ is the frustration parameter. We used the original Wang-Landau sampling and the standard Metropolis algorithm to confront past results of this model obtained by the effective-field theory (EFT) for the cubic lattice. Our numerical results suggest that the predictions of the EFT are in general qualitatively correct, but the low-temperature reentrant behavior, observed in the frontier separating the ferromagnetic and the colinear order, is an artifact of the EFT approach and should disappear when we consider Monte Carlo simulations of the model. In addition, our results indicate that the continuous phase transition between the Ferromagnetic and the Paramagnetic phases, that occurs for $0.0 \leq \alpha < 0.25$, belongs to the universality class of the three-dimensional pure Ising Model.

\vspace{0.5cm}


\end{abstract}

\maketitle

\section{Introduction\protect\nolinebreak}
Some magnetic compounds like $Eu_{x}Sr_{1-x}S$ \cite{comp1,comp2} and $Fe_{x}Zn_{1-x}F_{2}$ \cite{comp3} present more than one low-temperature magnetic ordering, depending on its parameters like the strength of the interactions and the concentration of magnetic ions $x$. They are well-described by models which consider competition of nearest-neighbor and next-nearest-neighbor interactions. The simplest model which may describe such compounds is  represented by the following Hamiltonian,

\begin{equation} \label{1}
\mathcal{H} = -J_{1} \sum_{nn} \sigma_{i}\sigma_{j} + J_{2} \sum_{nnn} \sigma_{l}\sigma_{m} ~,
\end{equation}%
where $\sigma_{i}= \pm 1$ are Ising spins ($i=1,\ldots,N$) and $N$ is the total number of spins. The first summation represents the exchange ferromagnetic interactions ($J_{1}>0$) between nearest-neighbor pairs of spins and the second term stands for the antiferromagnetic next-nearest-neighbor interactions ($J_{2} > 0$). The model with $J_{2} \leq 0$ is well understood and establishes the Ising second-order universality  class \cite{baxter}. Nevertheless, this  model  has attracted a lot of interest in the past, especially when implemented in the square lattice \cite{r4,r5,r6,r7,r8,r9,r10,r11,r12,r13,r14,r15,r16,r17,r18,r19,r20,r21,r22,r23,r24,r25,r26,r27,r28}. For this case, the zero-temperature magnetic ordering depends on  the value of the frustration parameter $\alpha = J_{2}/|J_{1}|$. For $\alpha < 1/2$, the order is ferromagnetic (F, $J_{1}>0$) or antiferromagnetic (AF, $J_{1}<0$), and for $\alpha > 1/2$, we have the collinear order, also called superantiferromagnetic order (SAF). For $1/2 <\alpha \leq 1$ there are controversial results about the nature of the order-disorder transition at finite temperatures. Recently, Kalz and Honecker \cite{kalz2} have concluded that Monte Carlo (MC) data obtained for large lattice sizes ($L=1000$, $2000$) indicate a clear picture only for $1/2 < \alpha < 1$, where a first-order phase transition scenario is established by the double-peaked structure of the energy histograms. Furthermore, for $\alpha \geq 1$ the phase transitions are of continuous type.

\begin{figure}[t]
\centering
\includegraphics[width=10.0cm,height=8.0cm]{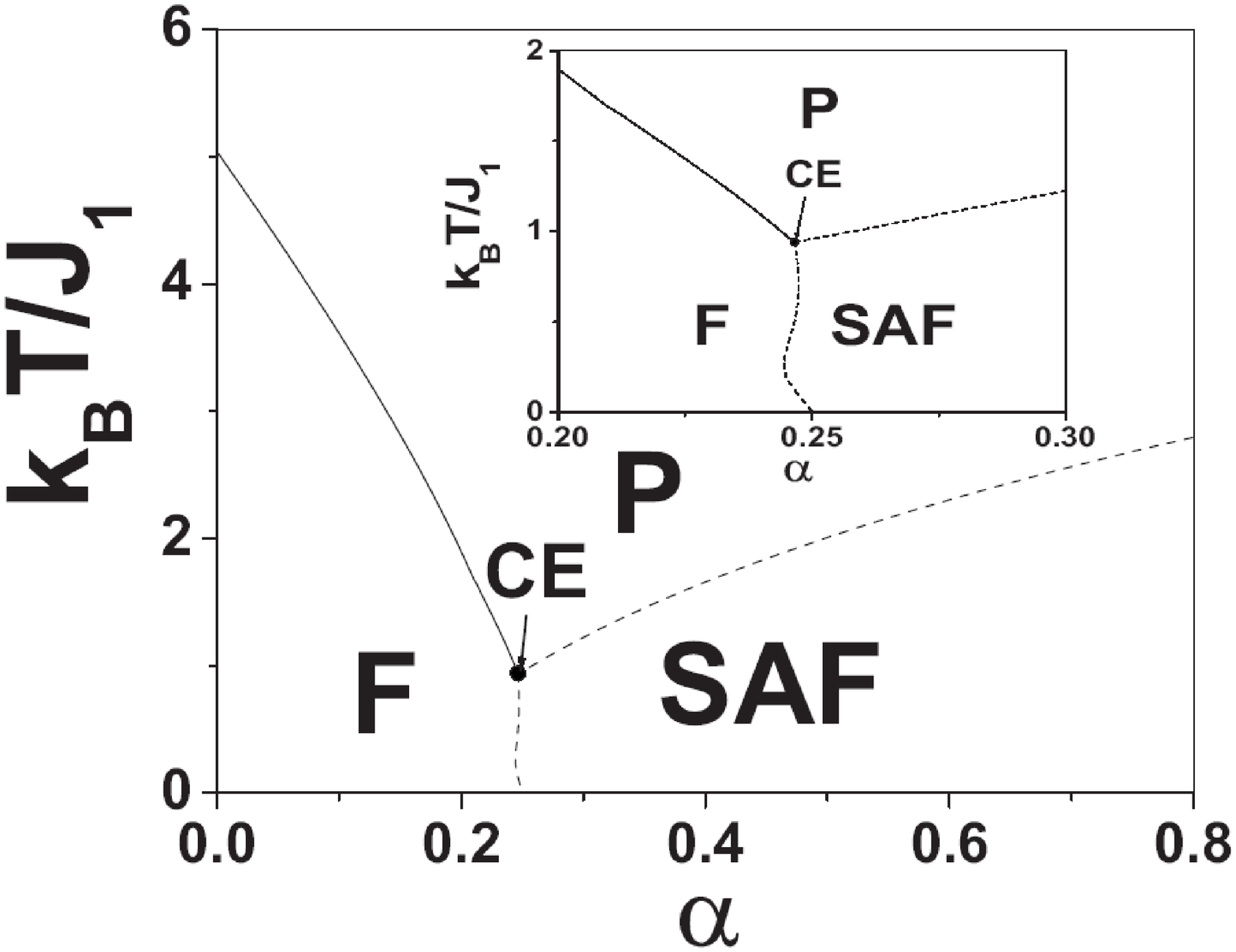}
\caption{Phase Diagram of the model described by the Hamiltonian in Eq. (\ref{1}), obtained from Ref. \cite{anjos}. The vertical axis represents the reduced temperature, and the horizontal one represents the frustration parameter ($\alpha=J_{2}/J_{1}$). Dashed and continuous lines represent first and second order critical frontiers, respectively. The inset shows better the reentrant form of the 
frontier separating the F and SAF orders. CE represents the critical end point. Figure obtained from Ref. \cite{anjos}.}
\label{fig1} 
\end{figure}

On the other hand, in three dimensions we do not have so many studies as in its two-dimensional counterpart and the situation is not so clear. A study using the Cluster Variational Method \cite{cirillo} has shown that the model has a first-order transition line separating the SAF phase from the disordered Paramagnetic (P) phase, as well as from the SAF phase to the F or AF phase, with the presence of a critical end point (CE). In this case, this model has been previously applied to treat random surfaces \cite{cappi,karowski} and microemulsions \cite{gompper}, and also as a discretized string action \cite{cirillo2}. In addition, the 3D model has already been treated within an effective-field theory by dos Anjos et al. \cite{anjos}. In Fig. \ref{fig1}, we show the phase diagram of the model in the plane $k_{B}T/J_{1}$ versus $\alpha$, obtained in \cite{anjos} by using an effective-field theory with a cluster of one central spin (EFT-1). At zero temperature, it can be exaclty determined two type of orderings separated by $\alpha = 1/4$. For $\alpha < 1/4$, the F order appears, whereas for $\alpha > 1/4$ the order is SAF. At finite temperatures these phases are separated by a first-order frontier, which presents a reentrant form as shown in the inset of Fig. \ref{fig1}. This frontier ends at a CE, where two order-disorder frontiers are also ending. The first one is of second-order type and separates the F and P phases for small values of $\alpha$, and the second one is of first-order type and separates the SAF and P phases, for higher values of $\alpha$. 

In the present work we study the model of Eq. (\ref{1}) defined on the cubic lattice with periodic boundary conditions, with $J_{1}>0$ and $J_{2}>0$. In the approximative methods like  the EFT-1 considered in \cite{anjos}, the authors used a decoupling procedure, which ignores all high-order correlations so as to approach the unmanageable expressions of all boundary spin-spin correlation functions. Although this analytical treatment improves the mean-field approach, which is insensitive to frustration, accuracy and qualitative aspects can be lost in determining the critical temperatures and the nature of the phase transitions \cite{yuksel1,yuksel2,akinci}. In order to study the qualitative aspects of the phase diagrams, we need to use powerful Monte Carlo techniques, which allow us to construct the canonical probability distribution function (CPDF) ($P(E,T) \sim \exp(-\beta E)$), for a given temperature and lattice size. Accordingly, at a critical temperature the $CPDF$ will show a double-peaked form for a first-order phase transition, or a single-peaked form for a second-order one. So, the original Wang-Landau sampling algorithm (WLS) is a suitable MC method to obtain the CPDF from the density of states $g(E)$ \cite{wlandau,wlandau2}.


\section{Methodology\protect\nolinebreak}

One of the advantages of the Wang-Landau method is that  we can directly construct  the density of states $g(E,T)$ through which the canonical partition function is achieved. Thus,  all the thermodynamic variables can be plotted as functions of temperature (free energy, heat capacity, etc). Furthermore,  the Metropolis algorithm will get trapped in states of local energy minima at low temperatures \cite{okamoto}, especially in frustated models. For instance, conventional simulations in the canonical ensemble would not be efficient  in the region close to $\alpha=0.25$, where the system is in a highly frustrated zone (see Fig. \ref{fig1}). Nevertheless, the original WLS does not present accuracy problems \cite{capa1}, and in this case it does not affect the results. However, another problem appears when large lattice sizes are needed. In this case, we require to divide the relevant energy range into fixed windows, then we have to join the windows after convergence is reached. Consequently, the resulting density of states and the associated thermodynamic functions suffer from boundary effects. This undesirable effect  becomes more conspicuous for the obtention of $g(E,M)$, which is necessary to calculate the CPDF, including the order parameter $P(E,M,T)$. In this case, it is necessary to perform a two-dimensional random walk in a relevant $(E,M)$ space. In most cases, this relevant $(E,M)$ space needs to be divided into surfaces to reach convergence, but after we have to join these surfaces, and the resulting function $P(E,M,T)$ will present small discontinuities.  

The general source of these difficulties seems to be due to the difficulty in matching surfaces at the boundaries rather than curves as in one-dimensional random walks \cite{wlandau3}. To overcome this problem, Cunha-Netto et al. proposed the  WLS with adaptive windows \cite{capa2}, where instead of defining fixed energy windows, the boundary positions depend on the set of energy values for which the histogram is flat at a given stage of the simulation. So, errors that may arise near the border of a given window are corrected in subsequent stages, for which the border positions are shifted. Nevertheless, it improves the quality of the results of the $WLS$ with fixed windows  at the expense of computational cost. The $WLS$ algorithm with adaptive windows considerably increases the computational time with the rise of the system size, and seems not to be able to be parallelized. Therefore, in this paper we use the multi-range original $WLS$ algorithm with fixed windows, which does not affect qualitative results as will be shown.

We used the original Wang-Landau algorithm in order to get the corresponding logarithm  of the density of states $\log g(E)$ for the model defined by Eq. (\ref{1}). Consequently, we can calculate the mean energy $E$ and the specific heat $C$, and therefrom the CPDF $P(E)$. These are our least necessary tools to do a qualitative analysis of the criticality of the system. In order to get $\log g(E)$, the minimum $E_{\min}$ and maximum $E_{\max}$ energies of the system are needed, for a given value of $\alpha$ and $L$. Then we also need to  number every discrete energy value $E_{j}$ between them  to define  an integer array $H(E_{j})$ and a real array $g(E_{j})$ as useful histograms for the algorithm. Initially, the $g(E)$ is unknown, so all bins in the array are set to unity. Since the typical range of $g(E)$ is of high orders of magnitude, it is common to store $\log g(E)$. In addition, a visit histogram $H(E)$ is maintained. 

Initially, all bins have zero visits for both $\log g(E)$ and $H(E)$. The bins are then filled over the course of a MC simulation, and the moves (spin flips) are accepted if $p < \min\left\{1, \frac{g(E)}{g(E')} \right\},$ where $p$ is a random number uniformly distributed in the range $[0,1]$, and $E$ and $E'$ are the energies of the current move and of the proposed one, respectively. After the move is accepted or rejected, the  histogram $H(E)$ is incremented by one and the density of states' histogram $g(E)$ is multiplied by a constant factor $f$ [$g(E) \rightarrow g(E)\times f$], where the initial choice is $f = e \simeq 2.72$. An accurate estimate of $g(E)$ is reached if the  histogram $H(E)$ becomes flat. 

At this step the  histogram $H(E)$ is set to zero and the modification factor $f$ is reduced such that $ f_{i+1} \rightarrow \sqrt[n]{f_{i}}$. This process is repeated until $f_{i}$ be close to 1, so we repeat it until $i=14$, using $n=4$ to accelerate the process. However, the repetition of the above simulation  suffers from the shortcoming that very large entries need to be stored in $g(E)$. As mentioned before, in order to avoid this problem, the quantity $ \log g(E) \rightarrow  \log  g(E)  + \log  f$ is evaluated. The modification factor is then now updated as $\log(f_{i+1}) \rightarrow (1/n)\log(f_{i})$. 

The adopted flatness criterion  was $H(E_{j}) > 0.8\left\langle H(E)\right\rangle$, $ \forall$ $j$. However, for the present model, it is difficult to satisfy it  around $\alpha=0.25$, due to frustration. So, for a given $f_{i}$ we stop the process after a maximum number of Monte Carlo moves ($M_{\max}$).   On the other hand, it  is important to mention that it is not necessary to use the entire energy interval $[E_{\min}\ldots E_{\max}]$ of the system to get the relevant information of the criticality. Thus, we need just to obtain the density of states for the relevant energy subspace $[E_{1},E_{2}]$ in order to calculate the thermodynamic quantities throughout the temperature range of our interest. For our model, and for a given lattice size $L$, the number of energy bins are considerably increased for some values of $\alpha$, and we have to apply a multi-range Wang-Landau algorithm with fixed windows even for the relevant energy subspace $[E_{1},E_{2}]$. Otherwise, the flatness criterion will never be satisfied.


\section{Results and Discussion}

\begin{figure}[t]
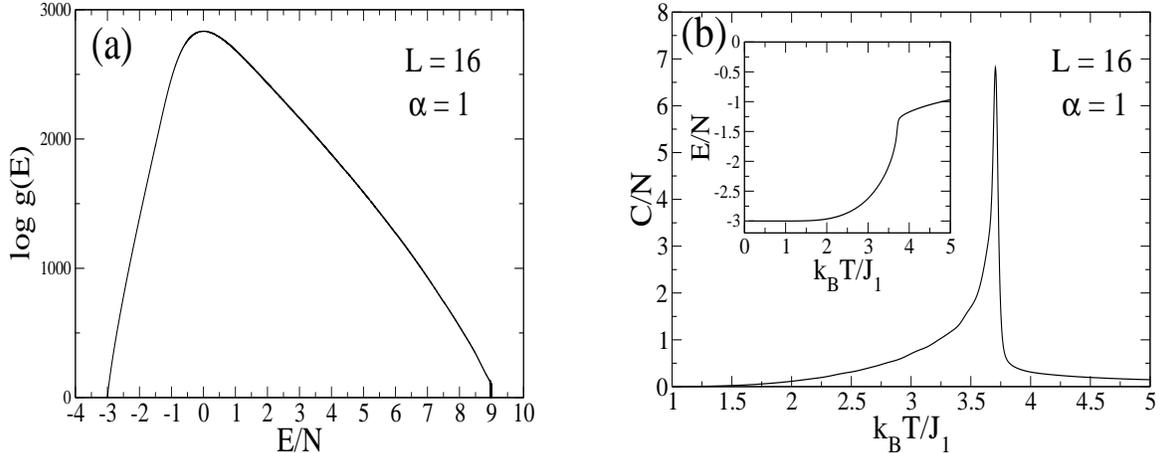

\centering
\includegraphics[width=7.0cm,height=6.0cm]{figure2a.eps}
\hspace{1.0cm}
\includegraphics[width=7.0cm,height=6.0cm]{figure2b.eps}
\caption{(a) Logarithm of the density of states for $\alpha=1.0$ and lattice size $L=16$, obtained by the original WLS algorithm applied for the whole energy range, without  dividing it by windows. (b) Specific heat associated with the energy shown in the inset, which was obtained from the density $g(E)$ exhibited (a).}
\label{fig2} 
\end{figure}

\begin{figure}[tb]
\centering
\vspace{0.4cm}
\includegraphics[width=8.0cm,height=6.0cm]{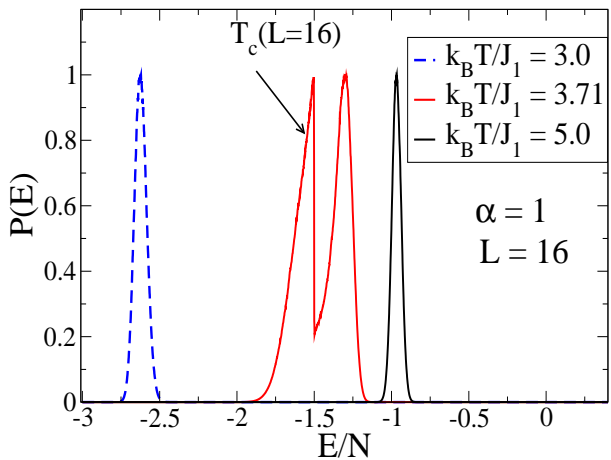}
\caption{CPDF for $\alpha=1.0$ and size $L=16$, at three different temperatures. This figure clearly suggests a first-order phase transition due to the double-peaked structure of the CPDF at the pseudo-critical temperature for this lattice size, $T_{c}(L=16) \approx 3.71$ (in units of $J_{1}/K_{B}$).}
\label{fig3} 
\end{figure}

We study the model defined in Eq. (\ref{1}) by performing the WLS and the Metropolis algorithm for $0.0 \leq \alpha \leq 1.0$, for the cubic lattice with $N=L\times L \times L$ sites. We choose $L=16$, because for $L>16$ the number of energy bins are considerably increased for certain values of $\alpha$. Thus, too many windows would be necessary to apply the  multi-range WLS algorithm, which would also increase the computational cost. In Fig. \ref{fig2} (a) it is exhibited the logarithm of the density of states $\log g(E)$ for $\alpha=1.0$, for the entire energy space. This is an  asymmetric function in $E$, in contrast to that of the simplest spin-1/2 Ising model. Fig. \ref{fig2} (b) shows the corresponding mean energy and the specific heat versus temperature obtained from $g(E)$.

\begin{figure}[tb]
\centering
\vspace{0.4cm}
\includegraphics[width=8.0cm,height=6.0cm]{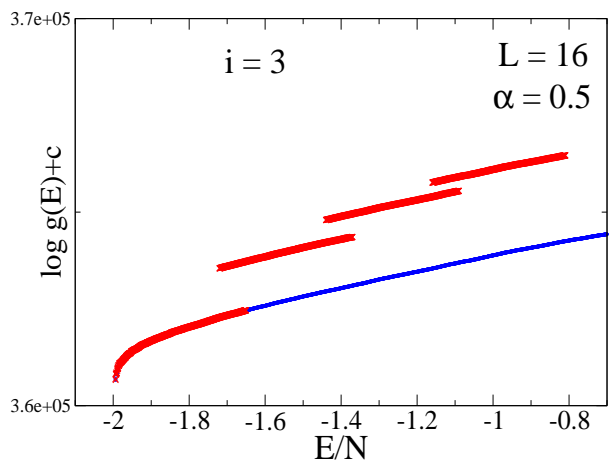}
\caption{Multi-range Wang-Landau results for the relative logarithm of the density of states, for $f_{i}=(\sqrt[4]{e})^{i}$, where $i=3$. The blue curve represents the overlapped windows.   }
\label{fig4} 
\end{figure}

\begin{figure}[t]
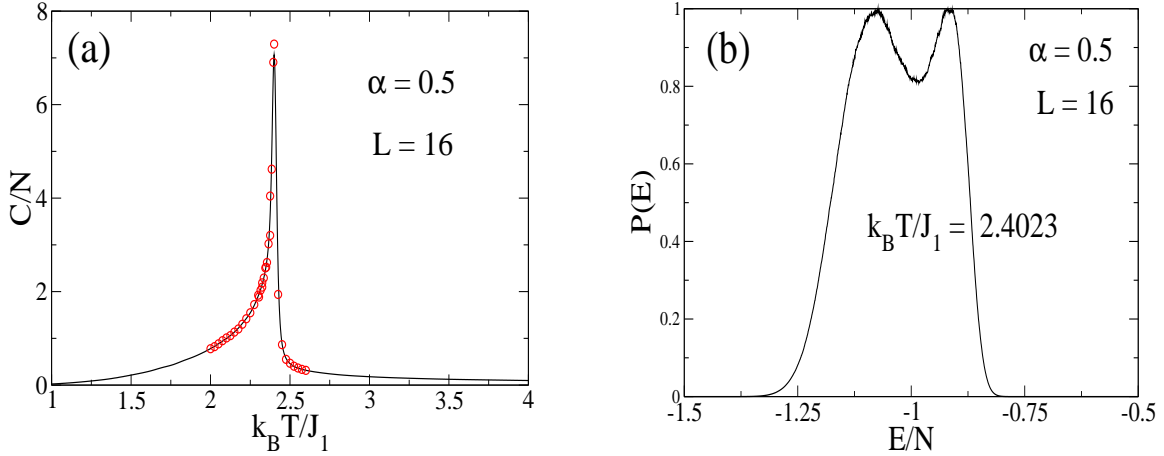

\centering
\includegraphics[width=7.0cm,height=6.0cm]{figure5a.eps}
\hspace{1.0cm}
\includegraphics[width=7.0cm,height=6.0cm]{figure5b.eps}
\caption{(a) Specific heat for $\alpha=0.5$ and $L=16$. The continuous line corresponds to WLS simulations, whereas red points to traditional Metropolis ones. (b) CPDF at the  pseudo-critical temperature $T_{c}(L=16) \approx 2.4$ (in units of $J_{1}/K_{B}$). The double-peaked structure of the CPDF suggests a first-order phase transition for the present size.}
\label{fig5} 
\end{figure}

\begin{figure}[tb]
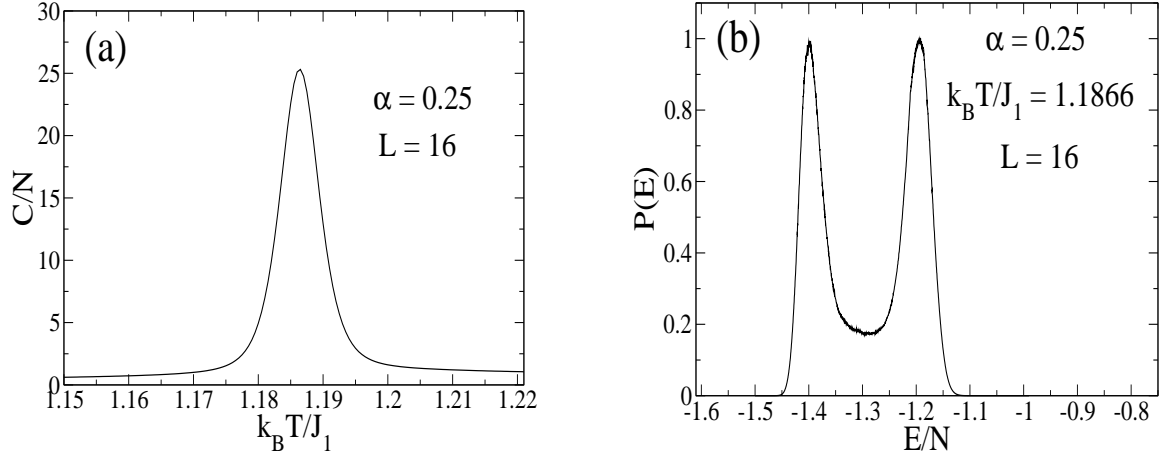

\centering
\vspace{0.7cm}
\includegraphics[width=7.0cm,height=6.0cm]{figure6a.eps}
\hspace{1.0cm}
\includegraphics[width=7.0cm,height=6.0cm]{figure6b.eps}
\caption{(a) Specific heat for $\alpha=0.25$ and $L=16$. (b) CPDF at the corresponding  pseudo-critical temperature $T_{c}(L=16) \approx 1.18$ (in units of $J_{1}/K_{B}$), showing a double-peaked structure.}
\label{fig6} 
\end{figure}

In Fig. \ref{fig3} we show the CPDF for $\alpha=1.0$ and three different temperatures obtained by the WLS. A double-peaked structure appears at the estimated pseudo-critical temperature $T_{c}(L=16)\approx 3.71$ (in units of $K_{B}/J_{1}$), suggesting a first-order phase transition. In Fig. \ref{fig4} we exhibit the energy range used to perform the WLS process for $\alpha=0.5$ and for a given step of the algorithm ($i=3$). We can see the results for the four overlapped windows in which the selected energy range was divided. Then we meet the curves of the four windows into one curve to get the logarithm of the density of states plus a constant. Accordingly, in Fig. \ref{fig5} we show the relevant results for $\alpha=0.5$. In Fig. \ref{fig5} (a) we exhibit the results for the specific heat, for which both the WLS and the traditional Metropolis simulations are in agreement. In addition, the double-peaked structure of the CPDF indicates the occurrence of a first-order transition at the estimated pseudo-critical temperature $T_{c}(L=16)\approx 2.4$ (in units of $K_{B}/J_{1}$, see Fig. \ref{fig5} (b)). For $\alpha=0.25$, the $F$ and SAF orders coexist at $T=0$. So, WLS results at finite temperatures present a first-order phase transition at the specific heat peak as shown in Figs. \ref{fig6} (a) and \ref{fig6} (b). It is necessary to point out that for this value of $\alpha$ the system is highly frustrated, and the traditional Metropolis simulations are not suitable to equilibrate the system, specially at low temperatures. Nonetheless, by the knowledge of the density of states, obtained by WLS, one may overcome the limitations of the traditional Monte Carlo technique.

\begin{figure}[tb]
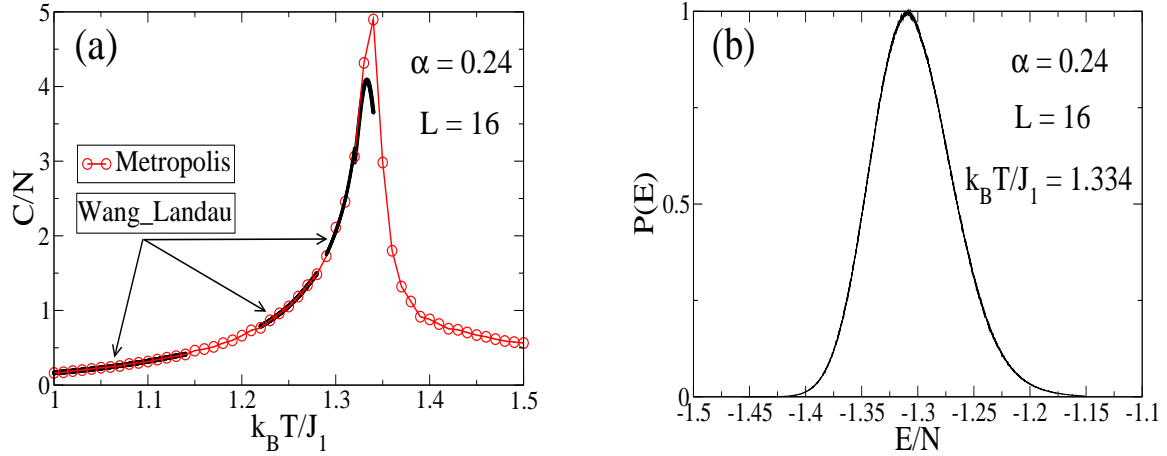

\centering
\vspace{0.4cm}
\includegraphics[width=7.0cm,height=6.0cm]{figure7a.eps}
\hspace{1.0cm}
\includegraphics[width=7.0cm,height=6.0cm]{figure7b.eps}
\caption{(a) Specific heat for $\alpha=0.24$ and $L=16$. The black lines correspond to WLS results for different energy ranges, whereas the red circles stand for traditional Metropolis results. (b) CPDF at the corresponding pseudo-critical temperature $T_{c}(L=16) \approx 1.33$ (in units of $J_{1}/K_{B}$) for the specific heat peak obtained by WLS. The single-peaked structure of the CPDF suggests a second-order phase transition for the present size.}
\label{fig7} 
\end{figure}

\begin{figure}[t]
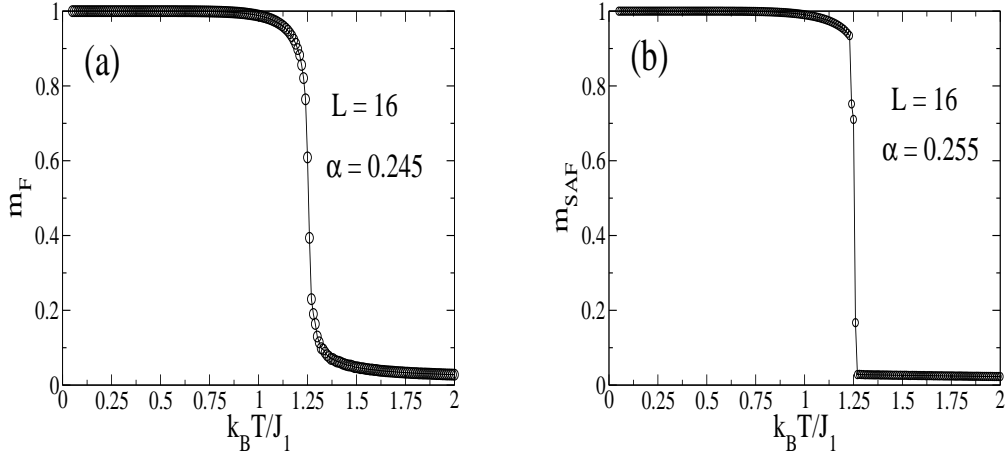

\centering
\includegraphics[width=6.0cm,height=6.0cm]{figure8a.eps}
\hspace{1.0cm}
\includegraphics[width=6.0cm,height=6.0cm]{figure8b.eps}
\caption{Relevant order parameters for two values of $\alpha$ around $0.25$ obtained by the Metropolis algorithm. (a) Ferromagnetic order parameter showing a continuous phase transition. (b) Superantiferromagnetic order parameter showing a discontinuous phase transition. No reentrant behavior is suggested.}
\label{fig8} 
\end{figure}

\begin{figure}[t]
\centering
\vspace{0.4cm}
\includegraphics[width=7.0cm,height=7.0cm]{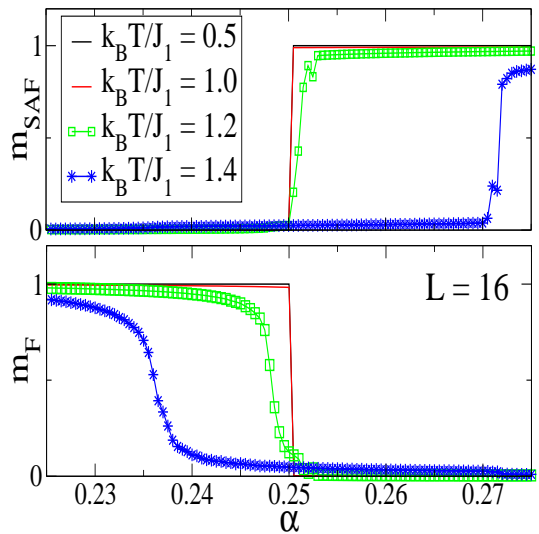}
\caption{Traditional Metropolis results for the order parameters versus $\alpha$, for $L=16$ and different temperatures.}
\label{fig9} 
\end{figure}

In order to verify whether a reentrant behavior occurs for the F-SAF frontier, as shown in Fig. \ref{fig1}, we need to explore values of $\alpha$ around $0.25$. In fact, the reentrant curve (obtained by EFT-1) occurs in the range $ 0.243 < \alpha < 0.25$. Nonetheless, we were not able to perform Wang-Landau simulations closer than $\alpha=0.24$. The reason is that even for $\alpha=0.24$ a relevant energy subspace requires many energy bins, even for single temperature calculations. Thus, we have built the specific heat curve by sections with the WLS method. In Fig. \ref{fig7} (a) we present these sections, and we can see that the WLS results agree with the Metropolis' simulations. At the specific heat peak, the corresponding CPDF shows a second-order phase transition in Fig. \ref{fig7} (b), because a single-peaked structure appears. To study the region close to $\alpha=0.25$, we have simulated the model for two values of $\alpha$, namely $\alpha=0.245$ and $\alpha=0.255$, by using the Metropolis algorithm. The results for the Ferromagnetic and Superantiferromagnetic order parameters $m_{{\rm F}}$ and $m_{{\rm SAF}}$, respectively, are exhibited in Fig. \ref{fig8}. In Fig. \ref{fig8} (a) we can see a second-order phase transition, whereas in Fig. \ref{fig8} (b) a first-order discontinuous transition takes place. These curves do not show more than one critical temperature, which suggests that there is no reentrance on the F-SAF frontier. Consequently, we might infere that the reentrance appeared in Fig. \ref{fig1} should be an artifact of the EFT-1 approach \cite{anjos}. On the other hand, we may approximately locate empirically the end of the F-SAF frontier by observing the behavior of the order parameters as functions of $\alpha$, as exhibited in Fig. \ref{fig9}. So, by the aid of this figure we  estimate the location of the CE around $T_{CE}= 1.0$ (in units of $J_{1}/k_{B}$) and $\alpha_{CE}=0.25$.

\begin{figure}[t]
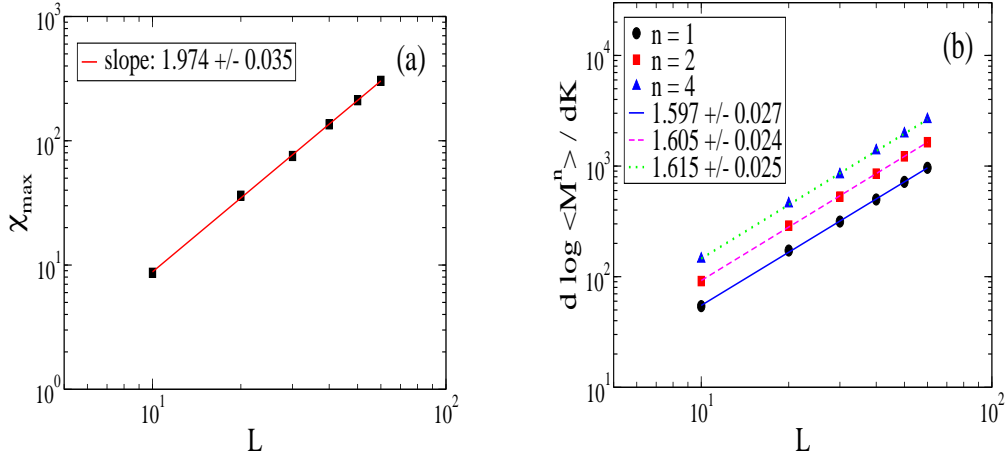

\centering
\includegraphics[width=6.0cm,height=6.0cm]{figure10a.eps}
\hspace{1.0cm}
\includegraphics[width=6.0cm,height=6.0cm]{figure10b.eps}
\caption{(a) The maxima of the susceptibility of the ferromagnetic order parameter versus the lattice size $L$ in the log-log scale, for $\alpha=0.1$. Fitting data, we can estimate the critical ratio $\gamma/\nu$ based on Eq. (\ref{2}), which give us $\approx 1.96$, that is equal to the 3D Ising model value within error bars. (b) The logarithmic derivatives of powers $n=1, 2$ and $4$ of the order parameter, defined by Eq. (\ref{3}), versus lattice size $L$ in the log-log scale, also for $\alpha=0.1$. Fitting data, we can estimate the critical ratio $1/\nu$, that is equal to the 3D Ising model value $\approx 1.6$ within error bars. Both results were obtained by the Metropolis algorithm, and the errors bars were estimated from the data fit.}
\label{fig10} 
\end{figure}

For $0.0 < \alpha < 0.25$ there is a second-order F-P critical frontier. Throughout this frontier, the three-dimensional Ising model universality class \cite{massimo,hasenbusch} seems to be unaffected by $\alpha$. In that range of $\alpha$, there are no problems to equilibrate the system, thus we have investigated the model by using the Metropolis algorithm for larger lattice sizes, up to $L=60$. For instance, Fig. \ref{fig10} (a) shows, in the log-log scale, the maximum of the susceptibility of the ferromagnetic order parameter versus the lattice size $L$, for $\alpha=0.1$. One can estimate the critical exponent ratio $\gamma/\nu$ based on the finite-size scaling equation
\begin{equation} \label{2}
\chi_{{\rm max}} \sim L^{\gamma/\nu} ~,
\end{equation} 

\noindent
that is valid in the vicinity of the phase transition. Thus, fitting data, we obtain the numerical estimate $\gamma/\nu \approx 1.96$ [see Fig. \ref{fig10} (a)], which is in agreement with the 3D Ising model universality class \cite{massimo,hasenbusch,landau91}. In addition, the estimation of the exponent $\nu$ is carried out by analyzing the divergence of the logarithmic derivatives of any power $n$ of the order parameter, defined as \cite{landau91,fytas2008}
\begin{equation} \label{3}
\frac{\partial{\ln \langle M^{n}\rangle}}{\partial{K}} = \frac{\langle M^{n}\,E\rangle}{\langle M^{n}\rangle} - \langle E\rangle ~,
\end{equation} 
\noindent
where $K=1/T$. As it is well known \cite{landau91}, the corresponding maxima scale with the system size as $\sim L^{1/\nu}$. In Fig. \ref{fig10} (b) we exhibit the size dependence of the first-, second- and fourth-order maxima of the average logarithm derivatives for $\alpha=0.1$. Fitting data, we obtained $1/\nu \approx 1.6$, which is also in agreement with the 3D Ising model universality class \cite{massimo,hasenbusch,landau91}. This same analysis was performed for other values of $\alpha$ in the range $0.0 < \alpha < 0.25$, where the transition is continuous, and we found the same critical exponents, considering the error bars, which confirms that the F-P frontier is universal, i.e., the disorder does not affect the universality  of the continuous phase transition of the three-dimensional Ising model. We have obtained all the results of Fig. \ref{fig10} by using the Metropolis algorithm, and the errors bars were estimated from the data fit.

\begin{figure}[t]
\centering
\includegraphics[width=7.0cm,height=6.0cm]{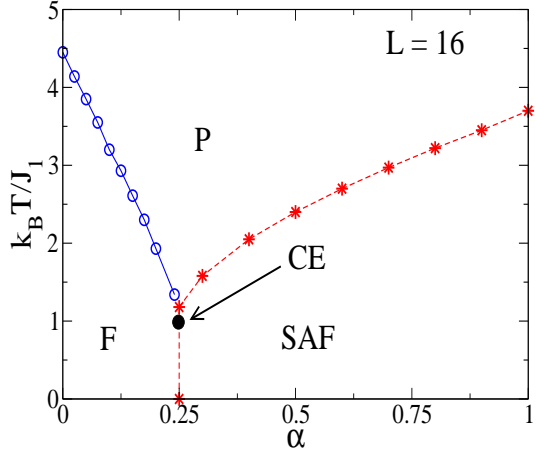}
\caption{Phase diagram of the model defined in Eq.(1) for $L=16$. The circles are the numerical estimates for the transition temperatures $T_{c}(L=16)$ of the second-order type, whereas the stars stand for first-order transition points. The location of the critical end point is uncertain, so it must be close to the end of the arrow. The star at $T=0$ and $\alpha=0.25$ is exactly located. All lines are just guides to the eye, and the error bars are smaller than data points, as discussed in the text.}
\label{fig11} 
\end{figure}

To summarize the results of the paper, we exhibit in Fig. \ref{fig11} the phase diagram of the present model in the plane $k_{B}T/J_{1}$ versus $\alpha$, for $L=16$. As discussed before, for $0.0 < \alpha < 0.25$ the phase transition between the Ferromagnetic (F) and the Paramagnetic (P) phases is of continuous type. In this case, for each value of $\alpha$ the pseudo-critical temperatures $T_{c}(L=16)$ were identified by the position of susceptibility (and specific heat) peak for $L=16$, obtained by the Metropolis algorithm. On the other hand, for $0.25 < \alpha < 1.0$ the system undergoes a first-order phase transition between the Superantiferromagnetic (SAF) and the Paramagnetic (P) phases. In this case, for each value of $\alpha$ we have analyzed the energy CPDF's for $L=16$ by using the Wang-Landau sampling. Thus, the transition points $T_{c}(L=16)$ were identified by the occurrence of a double-peaked structure of the CPDF. The processes for the identification of the two kinds of transitions allow us to estimate the error bars for the transition temperatures $T_{c}(L=16)$, but they are smaller than data points in Fig. \ref{fig11}. As discussed above, our results suggest the absence of reentrance, then the F-SAF frontier seems to be vertical. The three critical curves meet at a critical end point which must be around the end of the arrow shown in Fig. \ref{fig11}. Although the consideration of EFT approaches usually leads to some artificial results \cite{yuksel1,yuksel2,akinci}, in our case the results of Fig. \ref{fig11} show a phase diagram that agrees qualitatively with the phase diagram obtained by the EFT-1 \cite{anjos} (see Fig. \ref{fig1}), except by the reentrance that occurs in the last.

\section{Conclusions}
The phase diagram of the Ising model in the presence of nearest- and next-nearest-neighbor interactions, on a simple cubic lattice with $N=L \times L \times L$ sites, was studied by performing Monte Carlo simulations considering the original Wang-Landau sampling and the traditional Metropolis algorithm. The transition from the ordered ferromagnetic (F) phase to the disordered paramagnetic (P) phase is of second-order type, and the associated critical exponents belong to the 3D Ising model universality class. On the other hand, a first-order transition frontier is suggested from the superantiferromagnetic (SAF) phase to the P one, as well as from the SAF phase to the F one. The reentrance that appears in the F-SAF critical frontier obtained by an effective-field theory seems not to exist for the present formulation of the model. It suggests that this reentrance is a consequence of the limitations of the EFT approach. However, MC results give qualitatively the same phase diagrams as obtained by effective-field calculations. 

The Ising model with competing nearest- and next-nearest-neighbor interactions studied in this work can give a theoretical description of some magnetic compounds like $Eu_{x}Sr_{1-x}S$  and $Fe_{x}Zn_{1-x}F_{2}$ \cite{comp1,comp2,comp3}, that present more than one low-temperature magnetic ordering and different kinds of phase transitions \cite{lara,masrour,kaya,yuksel3,crok}. However, this kind of competition among positive (ferromagnetic) and negative (antiferromagnetic) interactions can be also useful to describe the dynamics of some social systems. For example, in opinion dynamics the presence of positive/negative interactions can models the agreement/disagreement among individuals, and a mean-field Ising-like universality class can be identified \cite{meu_celia}. In addition, the dynamics of cooperation/defection in evolutionary games can also be seen as a practical situation where positive and negative interactions occur. In fact, in some of these models second- and first-order phase transitions are present \cite{szolnoki1,szolnoki2,szolnoki3}. For both social systems, it can be interesting to analyze the effects of interactions among individuals in a given lattice (square, cubic) and their nearest and next-nearest neighbors. The presence of such interactions with competitive positive and negative signals can give some interesting results for the field of social dynamics.

\section*{Acknowledgements}

This work was partially supported by CNPq (Edital Universal), CAPES, FAPERJ and FAPEAM (UNIVERSAL AMAZONAS e Programa Primeiros Projetos - PPP) (Brazilian Research Agencies).

\end{document}